\documentclass{IEEEtran4PSCC}

\usepackage{cite}

\ifCLASSINFOpdf
   \usepackage[pdftex]{graphicx}
\else
   \usepackage[dvips]{graphicx}
\fi
\usepackage[cmex10]{amsmath}
\usepackage{algorithmic}

\usepackage{array}

\ifCLASSOPTIONcompsoc
 \usepackage[caption=false,font=normalsize,labelfont=sf,textfont=sf]{subfig}
\else
 \usepackage[caption=false,font=footnotesize]{subfig}
\fi
\usepackage{stfloats}
\usepackage{url}

\usepackage{mathrsfs}
\usepackage{amssymb,amsfonts}
\usepackage{color}
\usepackage[normalem]{ulem}
\usepackage{enumitem}
\usepackage[utf8]{inputenc}
\usepackage{textcomp}
\usepackage{eurosym}
\usepackage{accents}
\newcommand{\ubar}[1]{\underaccent{\bar}{#1}}
\usepackage{hyperref}
\usepackage{blindtext}
\usepackage{xcolor}
\usepackage[normalem]{ulem}
\usepackage{booktabs}
\usepackage{acronym}
\usepackage[mathscr]{eucal}
\usepackage{mathtools}
\usepackage{graphicx}
\usepackage{pifont}
\newcommand{\xmark}{\ding{56}}%
\usepackage{multirow}
\usepackage{nomencl}
\newcommand{\nomunit}[1]{%
\renewcommand{\nomentryend}{\hspace*{\fill}#1}}
\makenomenclature

\usepackage{etoolbox}
\renewcommand\nomgroup[1]{%
  \item[\bfseries
  \ifstrequal{#1}{I}{Indexes and Sets}{%
  \ifstrequal{#1}{P}{Parameters}{%
  \ifstrequal{#1}{V}{Variables}{}}}%
]}

\acrodef{dam}[DAM]{Day-Ahead Market}
\acrodef{idm}[IDM]{Intra-Day Market}
\acrodef{rtm}[RTM]{Real-Time Market}
\acrodef{stu}[STU]{Solar Thermal Unit}
\acrodef{csp}[CSP]{Concentrated Solar Plant}
\acrodef{ess}[ESS]{Energy Storage System}
\acrodef{bess}[BESS]{Battery Energy Storage System}
\acrodef{res}[RES]{Renewable Energy Source}
\acrodef{soc}[SoC]{State of Charge}
\acrodef{ndres}[ND-RES]{Non-dispatchable Renewable Energy Source}
\acrodef{dres}[D-RES]{Dispatchable Renewable Energy Source}
\acrodef{ro}[RO]{Robust Optimization}
\acrodef{sth}[S-TH]{Solar Thermal Unit}
\acrodef{vpp}[VPP]{Virtual Power Plant}
\acrodef{pv}[PV]{photovoltaic}
\acrodef{bes}[BES]{Battery Energy Storage}
\acrodef{phes}[PHES]{Pumped Hydro Energy Storage}
\acrodef{opf}[OPF]{Optimal Power Flow}
\acrodef{pcc}[PCC]{Point of Common Coupling}

\makeatletter
\let\old@ps@headings\ps@headings
\let\old@ps@IEEEtitlepagestyle\ps@IEEEtitlepagestyle
\def\psccfooter#1{%
    \def\ps@headings{%
        \old@ps@headings%
        \def\@oddfoot{\strut\hfill#1\hfill\strut}%
        \def\@evenfoot{\strut\hfill#1\hfill\strut}%
    }%
    \def\ps@IEEEtitlepagestyle{%
        \old@ps@IEEEtitlepagestyle%
        \def\@oddfoot{\strut\hfill#1\hfill\strut}%
        \def\@evenfoot{\strut\hfill#1\hfill\strut}%
    }%
    \ps@headings%
}
\makeatother

\begin{document}

\setlength{\abovedisplayskip}{0.5pt}
\setlength{\belowdisplayskip}{0.5pt}

\title{Optimal Participation of Heterogeneous, RES-based Virtual Power Plants in Energy Markets \vspace*{-0.25cm}}

\author{
\IEEEauthorblockN{Oluwaseun Oladimeji, {\'{A}}lvaro Ortega, Lukas Sigrist, Luis Rouco, Pedro S\'{a}nchez-Mart\'{i}n, Enrique Lobato}
\IEEEauthorblockA{Institute for Research in Technology, Comillas Pontifical University, Madrid, Spain}
\vspace*{-9mm}
}

\maketitle

\begin{abstract}
In this work, we present a detailed model of an \ac{res}-based \ac{vpp} that participates in \ac{dam} and \ac{idm} with dispatchable and non-dispatchable \acp{res} and flexible demand assets. We propose a demand model with bi-level flexibility which are associated with the market sessions plus an improved solar thermal plant model with piece-wise linear formulation of efficiency. A network-constrained unit commitment model is used by the \ac{vpp} to submit \ac{dam} auctions and consequently participates in \ac{idm} to correct for deviations. Finally, we validate our model by assessing its operation on different weather conditions of uncertainty.
\end{abstract}

\begin{IEEEkeywords}
Day Ahead Market, Flexible Load, Intra-Day Market, Renewable Energy Sources, Solar Thermal Plants, Virtual Power Plant
\end{IEEEkeywords} 
\vspace*{-0.75cm}

\thanksto{\noindent This project has received funding from the European Union’s Horizon 2020 research and innovation programme under grant agreement No 883985 (POSYTYF project).}

\begin{small} 
\setlength{\nomitemsep}{0.05cm}

\nomenclature[I, 01]{$b \in \mathscr{B}/\mathscr{B}^m$}{Network Buses / Network buses Connected to Main Grid}
\nomenclature[I, 01]{$c \in \mathscr{C}/\mathscr{C}_b$}{\ac{dres} / \ac{dres} connected to bus $b$}
\nomenclature[I, 01]{$d \in \mathscr{D}/\mathscr{D}_b$}{Demand / Demand connected to bus $b$}
\nomenclature[I, 02]{$\ell \in \mathscr{L}$}{Network lines}
\nomenclature[I, 01]{$k \in \mathscr{K}$}{\ac{idm} sessions}
\nomenclature[I, 03]{$p \in \mathscr{P}$}{Demand profiles}
\nomenclature[I, 03]{$r \in \mathscr{R}/\mathscr{R}_b$}{\ac{ndres} / \ac{ndres} connected to bus $b$}
\nomenclature[I, 03]{$t \in \mathscr{T}$}{Time periods}
\nomenclature[I, 04]{$\theta \in \Theta/\Theta_b$}{\ac{stu} / \ac{stu} connected to bus $b$ \vspace{5pt}}

\nomenclature[P, 01]{$C_c^0/C_c^1$}{Shut-down/start-up cost of \acp{dres}\nomunit{[\euro]}}  
\nomenclature[P, 01]{$C_c^{\rm V}$}{Variable production cost of \acp{dres}\nomunit{[\euro/MWh]}}
\nomenclature[P, 03]{$\ubar{P}_c/\bar{P}_c$}{Min/Max power production of \acp{dres}\nomunit{[MW]}}

\nomenclature[P, 01]{$C_{d,p}$}{Cost of load profile $p$ of demand\nomunit{[\euro]}}
\nomenclature[P, 02]{$\ubar{E}_d$}{Min energy consumption of demand $d$ throughout the planning horizon\nomunit{[MWh]}}
\nomenclature[P, 03]{$P_{d,p,t}$}{Max hourly consumption of profile $p$ of demand $d$ \nomunit{[MW]}}
\nomenclature[P, 03]{$\ubar{P}_{d,t}/\bar{P}_{d,t}$}{Lower/upper bound of the power consumption of demand $d$ in time $t$\nomunit{[\%]}}
\nomenclature[P, 05]{$\ubar{R}_d/\bar{R}_d$}{Down/up ramping limit of demand $d$\nomunit{[MW/h]}}

\nomenclature[P, 03]{$\ubar{P}_{r,t}$}{Min production of \ac{ndres} in time $t$\nomunit{[MW]}}
\nomenclature[P, 04]{$\check{P}_{r,t}$}{Available production of \ac{ndres} in time $t$\nomunit{[MW]}}

\nomenclature[P, 02]{$K_\theta$}{\ac{stu} $\theta$ output multiplier at startup\nomunit{[$-$]}}
\nomenclature[P, 03]{$\ubar P_{\theta}/\bar P_{\theta}$}{Min/max production capacity of \ac{stu} $\theta$\nomunit{[MW]}}
\nomenclature[P, 04]{$\check{P}_{\theta,t}$}{Available power production of \ac{stu} in time  $t$\nomunit{[MW]}}
\nomenclature[P, 03]{$\ubar{P}_{\theta}^+/\bar{P}_{\theta}^+$}{Lower/upper bound of charging capacity of \ac{stu} storage $\theta$\nomunit{[MW]}}
\nomenclature[P, 03]{$\ubar{P}^-_{\theta}/\bar{P}^-_{\theta}$}{Lower/upper bound of discharging capacity of \ac{stu} storage $\theta$\nomunit{[MW]}}
\nomenclature[P, 07]{$\eta_{\theta}^+/\eta_{\theta}^-$}{Charging/discharging efficiency of \ac{stu} storage\nomunit{[\%]}}
\nomenclature[P, 07]{$\eta_{\theta}^{n}$}{Conversion factor between thermal and electrical power in the PB of \ac{stu} $\theta$ in segment $n\in\{1,2,3,4\} $ \nomunit{[\%]}}

\nomenclature[P, 05]{$\bar{P}_b^m$}{Maximum power that can be traded with the main grid at bus $b$ \nomunit{[MW]}}

\nomenclature[P, 07]{$\lambda^{\rm DA}_t$}{\ac{dam} price in time $t$\nomunit{[\euro/MWh]}}
\nomenclature[P, 07]{$\lambda^{\rm ID}_{k,t}$}{Price of \ac{idm} session $k$ in time  $t$\nomunit{[\euro/MWh]} \vspace{5pt}}
\nomenclature[P, 07]{$\Delta t$}{Duration of time periods\nomunit{[h, min]}}
\nomenclature[P, 06]{$T$}{Last period of schedule\nomunit{[$-$]}}

\nomenclature[V, 01]{$p_{c,t}$}{Power generation of \acp{dres} in time $t$\nomunit{[MW]}}

\nomenclature[V, 01]{$p_{d,t}$}{Power consumption of demand in time $t$\nomunit{[MW]}}
\nomenclature[V, 01]{$u_{d,p}$}{Binary variable to select demand profile\nomunit{[$0/1$]}}

\nomenclature[V, 01]{$p_{r,t}$}{Power generation of \ac{ndres} in time $t$\nomunit{[MW]}}

\nomenclature[V, 01]{$p_{\theta,t}$}{Electrical power generation of \ac{stu} in time $t$\nomunit{[MW]}}
\nomenclature[V, 01]{$p_{\theta,t}^+/p_{\theta,t}^-$}{Charging/discharging thermal power level of \ac{stu} storage $\theta$ in time  $t$\nomunit{[MW$_t$]}}
\nomenclature[V, 01]{$u_{\theta,t}^+$}{Binary variable to control \ac{stu} storage charging\nomunit{[$0/1$]}}
\nomenclature[V, 01]{$u_{\theta,t}$}{Binary variable to control \ac{stu} PB operation \nomunit{[$0/1$]}}
\nomenclature[V, 01]{$v_{\theta,t}^1$}{Binary variable to control startup of \ac{stu} PB  $\theta$ \nomunit{[$0/1$]}}
\nomenclature[V, 01]{$p_{\theta,t}^{\rm PB}$}{Thermal power delivered to the \ac{stu} power block \nomunit{[MW]}}
\nomenclature[V, 01]{$p_{\theta,t}^{\rm SF}$}{Thermal power generated by the solar field \nomunit{[MW]}}

\nomenclature[V, 01]{$p_{\ell,t}$}{Power flow through network of line $\ell$ in time $t$\nomunit{[MW]}}

\nomenclature[V, 01]{$p_{b,t}^m$}{Power scheduled to be bought from/sold to the \ac{dam} and \ac{idm} markets at bus $b$ in time  $t$\nomunit{[MW]}}

\nomenclature[V, 01]{$p_t^{\rm DA}$}{Total power traded in the \ac{dam} in time $t$\nomunit{[MW]}}
\nomenclature[V, 01]{$p_t^{\rm ID}$}{Total power traded in the \ac{idm} in time $t$\nomunit{[MW]}}

\printnomenclature[1.5cm]

\end{small} 

\section{Introduction} \label{sec: intro}
Nowadays, \acp{res} have emerged as a crucial part of modern power systems due mainly to their decreasing costs of operation and minimal carbon footprint. However, most \ac{res} technologies depend on sources of stochastic nature, and are non-dispatchable. Stochastic \acp{res} thus have inherent disadvantage when participating in energy markets, as they are susceptible to economic penalties and/or losses if they do not supply the energy scheduled in the corresponding market session \cite{pinson2017towards}. Additionally, they have relatively small sizes as single offering units compared with large conventional, synchronous plants.

The aggregation of \acp{res} as \acp{vpp} to provide a more controllable output is a promising solution to improve the competitiveness of stochastic \acp{res} in energy markets. A \ac{vpp} participating in electricity markets and comprising only wind and solar PV power plants was presented in \cite{koraki2017wind}. The uncertainties related to wind and irradiation were dealt with using appropriate forecasts and integrating the \ac{vpp} operation with the distribution network. In \cite{dominguez2014operation}, a \ac{vpp} consisting of only stochastic \acp{res} was analysed and energy storage systems were utilized in providing the flexibility required to reduce the impact of generation uncertainty. 


While models of the most common stochastic \ac{res} technologies such as wind and solar PV generation participating in energy markets are well established, appropriate modeling of other emerging technologies, e.g., \acf{csp} is still an open issue. For instance, two aspects need to be carefully considered when modelling \acp{csp}. These are integration of molten-salt thermal storage with the \acp{csp}, and non-linear conversion efficiencies from thermal to electrical energy. The latter depend on the level of thermal power injected into the power block of the plant. \acp{csp} with storage were modelled in \cite{sioshansi2010value} and \cite{He2016} but conversion efficiency was chosen as an average value in both studies. Choosing a single average efficiency value might appear as a good compromise \cite{rohani2017modelling}, but it does not correspond to actual operation of the \ac{csp}.

\acp{vpp} are also gradually including demands with flexibility provision capability in their portfolio \cite{liu2018optimal}. However, a main challenge with demand side management models is the absence of appropriate structures of incentives for consumers that provide such flexibility actions. 
Flexible and non-flexible loads with associated costs were presented in \cite{de2018local} but these incentive measures were not addressed. A price-based control for demand management was proposed in \cite{herter2007residential} while neglecting the parametrization of the comfort of end users. In \cite{hungerford2019value}, the behaviour of dispatchable loads on an aggregated scale was studied  but the end-goal in that work was decreasing the power system operation costs as well as reducing the need for conventional power plants.

In this regard, the contribution of this paper is a detailed model of heterogeneous, RES-based VPPs participating in energy markets. We propose:
\begin{enumerate}
    \item a demand model with bi-level flexibility associated with different energy market trading sessions. In contrast to other studies, the demand owner sets different profiles which the \ac{vpp} manager can choose from in \ac{dam} and allows tolerances around the chosen profile at \ac{idm};
    \item a detailed \ac{csp} model with storage capability. The model includes a linear formulation for the operation of \ac{csp} that addresses conversion from thermal to electrical energy using a piece-wise linear efficiency function; and
    \item a network-constrained unit commitment model used by the \ac{vpp} to submit \ac{dam} auctions and then subsequently participates in \ac{idm} sessions to correct for deviations of its \ac{ndres} forecasts. 
\end{enumerate} 
In this study, it is considered that the \ac{vpp} participates in the Spanish energy market because of its separation of \ac{dam} and \ac{idm}, and ease of entrance of \acp{res} \cite{ciarreta2011renewable}. However, no subsidies from the system or market operator are considered.


\section{Overview of Spanish Energy Market Structure} \label{sec: spanish mkt}
The wholesale electricity spot market in Spain is organised 
by the market operator (OMIE) \cite{omiees} and system operator (Red El{\'e}ctrica de España (REE))~\cite{redelectrica}. Two broad markets, the \ac{dam} and \ac{idm}, govern the day-to-day running of electricity provision and consumption with some additional arrangements associated with ensuring reliability and secure operation of the energy system~\cite{chaves2015spanish}.


After \ac{dam} clearance, 
\acp{idm} are instantiated to make adjustments to \ac{dam} cleared bids and correct infeasible schedules. These modifications or corrections can be due to unplanned shutdown of dispatchable sources of generation, changes in \ac{ndres} outputs, demand changes and/or line faults. \ac{idm} sessions are especially useful for balancing renewable generation bid deficits or surplus by giving market participants with RESs an avenue to update their availability when their submitted \ac{dam} bids are different from near real time realizations. There are seven \ac{idm} sessions; the first and second cover 24 hours of the operation day while the remaining five sessions cover a receding subset of the 24 hours \cite{chaves2015spanish}. 

A consortium of generation and demand assets (a \ac{vpp}), participating in the Spanish energy market can thus capture the operational flexibility which the different energy market segments offer in order to maximize its revenue.
Our work, while currently tailored to the Spanish market, can be applied in other markets of similar characteristics and proves the potential of RES-based \acp{vpp}.

\section{VPP Modelling} \label{sec: vpp model}
This section formulates and discusses the \ac{vpp} model proposed in this paper. The \ac{vpp} comprises \acfp{dres} (hydro and biomass), \acfp{ndres} (wind power plant, solar PV and solar thermal generation with storage capability) and flexible demands. These assets are distributed across the power network at different buses and connected to the main grid through one or more Points of Common Coupling (\acs{pcc}). The business model considered for the \ac{vpp} is maximization of its aggregated profit by optimally scheduling its generation and demand assets.

The formulation for each asset class is enumerated below. \ac{dres} are modeled like conventional power plants \cite{baringo2020virtual} with linearized operation costs of the dispatchable assets. Network constraints are formulated using DC power flow \cite{kargarian2016toward}. The objective function of the \ac{vpp}, and constraints for the \ac{ndres}s, flexible demand, \ac{stu} and energy balance at the PCCs are presented in the following subsections.

\subsection{Profit Maximization Objective}
Due to small volumes of energy traded in the \ac{idm} relative to \ac{dam} and modest price differences between these markets \cite{chaves2015spanish}, the objective functions in \ac{dam} and \ac{idm} are decoupled in this work.
Each \ac{idm} further has  associated constraints to cater for updates or changes in forecasts of stochastic sources.
In \ac{dam}, the objective function \eqref{obj:DAM_det} is the maximization of the obtainable profit by the \ac{vpp} assets calculated as the revenue from power trades minus cost of operating \ac{dres} and cost of selecting a particular load profile. For the different \ac{idm} sessions, the benefit \eqref{obj:IDM_det} is calculated over changes in traded power $\Delta p_{k,c,t}$ between: (i) \ac{dam} and first \ac{idm} trading period and (ii) other subsequent \ac{idm} sessions. Cost of choosing a specific load profile is not included while computing objective of \ac{idm} because this choice is previously made during \ac{dam} participation and must be accounted for only once. 
\begin{multline} \label{obj:DAM_det}
    \max_{\Xi^{\rm DAM}} 
    \sum_{t\in\mathscr{T}} 
    \left[ 
    \lambda^{\rm DA}_t p_t^{\rm DA} \Delta t - 
    \sum_{c\in\mathscr{C}} 
    \left( 
    C_c^{\rm V} p_{c,t} \Delta t + 
    c_{c,t}^0 + 
    c_{c,t}^1 
    \right) 
    \right] \\ 
    - \sum_{d \in \mathscr{D}} 
    \sum_{p \in \mathscr{P}} 
    C_{d,p}u_{d,p} \qquad \qquad \qquad \qquad \qquad \qquad \quad \vspace{-1mm}
\end{multline}
\begin{multline} \label{obj:IDM_det}
    \max_{\Xi^{\rm IDM}_k} 
    \sum_{t=\tau_k}^{\left| \mathscr{T} \right|} 
    \Bigg[
    \lambda^{\rm ID}_{k,t}  
    p_{k,t}^{\rm ID} \Delta t \\
    - \sum_{c\in\mathscr{C}} 
    \left( 
    C_c^{\rm V} \Delta p_{k,c,t} \Delta t + 
    c_{k,c,t}^0 + 
    c_{k,c,t}^1 
    \right) 
    \Bigg], 
    \quad 
    \forall k \in \mathscr{K}
\end{multline}

\subsection{Energy Balance}
Energy balance constraints common to both market stages are modeled in \eqref{cons:balance common} whereas those specific to \ac{dam} and \ac{idm} are formulated in \eqref{cons:balance_DAM} and \eqref{cons:balance_IDM} respectively.
Nodal analysis is utilized, implying that at each bus of the network, sum of inflows and outflows must be equal. Equation \eqref{cons:balance_mg} gives energy balance at the \ac{pcc} with the main grid while \eqref{cons:balance_no_mg} is the balance for all other buses in the \ac{vpp} network at every time period \cite{van2014dc}. The difference between both equations is the presence of $p_{b,t}^m$ at the main grid representing scheduled power to be traded with other market participants. This power available for trading (buy or sell) is set within prespecified bounds in \eqref{cons: trade_bound}.
\begin{subequations} \label{cons:balance common}
\begin{multline} \label{cons:balance_mg}
    \sum_{c\in\mathscr{C}_b} p_{c,t} \!+\!\! 
    \sum_{r\in\mathscr{R}_b} p_{r,t} \!+\!\! 
    \sum_{\theta\in\Theta_b} p_{\theta,t} \!-\!\! \!\!\!
    \sum_{\ell | i(\ell)=b} \!\!p_{\ell,t} +\! \!\!\!
    \sum_{\ell | j(\ell)=b} \!\!\!p_{\ell,t} \\
    = p_{b,t}^m \!+\!\! 
    \sum_{d\in\mathscr{D}_b} p_{d,t}~, 
    \qquad 
    \forall b \in \mathscr{B}^m, 
    \forall t \in \mathscr{T} 
\end{multline}
\begin{multline} \label{cons:balance_no_mg}
    \sum_{c\in\mathscr{C}_b} p_{c,t} \!\!+\!\! 
    \sum_{r\in\mathscr{R}_b} p_{r,t} \!\!+\!\! 
    \sum_{\theta\in\Theta_b} p_{\theta,t} -\!\! \!\!
    \sum_{\ell | i(\ell)=b} \!\!p_{\ell,t} +\!\!  \!\!
    \sum_{\ell | j(\ell)=b} \!\!p_{\ell,t} \\ 
    = \sum_{d\in\mathscr{D}_b} p_{d,t}~,
    \qquad 
    \forall b \in \mathscr{B} \setminus \mathscr{B}^m, 
    \forall t\in \mathscr{T}
\end{multline} \vspace{1ex}
\begin{IEEEeqnarray}{lr}
    -\bar{P}_b^m 
    \leq p_{b,t}^m 
    \leq \bar{P}_b^m ~,
    & \qquad \qquad \;\;
    \forall b \in \mathscr{B}^m, 
    \forall t \in \mathscr{T}  \label{cons: trade_bound}
\end{IEEEeqnarray} \vspace{0.0em}
\end{subequations} 
\vspace{-2mm}
\subsubsection{DAM Formulation} 
Equation \eqref{cons:balance_pDA2} ensures that summation of traded power at all buses connected to the main grid is equivalent to the total power available for trading by \ac{vpp} whereas \eqref{cons:balance_pDA1} relaxes this available power at each time period.
\vspace{-2mm}
\begin{subequations} \label{cons:balance_DAM}
\begin{IEEEeqnarray}{lr}
    p_t^{\rm DA} = \!\!
    \sum_{b\in \mathscr{B}^m} p_{b,t}^m ~, 
    & \qquad \qquad \qquad \qquad \qquad 
    \forall t \in \mathscr{T}  \label{cons:balance_pDA2}
\end{IEEEeqnarray}
\begin{multline} \label{cons:balance_pDA1}
    -\left( 
    \sum_{d\in\mathscr{D}} P_{d,p,t} \!+\! 
    \sum_{\theta\in\Theta}\!\bar{P}_{\theta}^ + 
    \right) 
    \leq 
    p_t^{\rm DA} 
    \\ \leq 
    \sum_{c\in\mathscr{C}} \bar{P}_{c} \!+\! 
    \sum_{r\in\mathscr{R}} \!\check{P}_{r,t} \!+\!  
    \sum_{\theta\in\Theta}\!\bar{P}_{\theta}~,
    \qquad 
    \forall p \in \mathscr{P},
    \forall t \in \mathscr{T} 
\end{multline} 
\end{subequations} \vspace{0mm}

\subsubsection{IDM Formulation}
For \ac{idm} sessions, the bounds on the traded power are modified such that the load profile that was chosen in the \ac{dam} is taken into account as formulated in \eqref{cons: energy_bal_idm1}. Note that the \ac{idm} offers/bids do not substitute those submitted in the \ac{dam}, but rather, they are \textit{adjustments} of the \ac{dam} offers/bids as reflected in \eqref{cons: energy_bal_idm3}. The rationale behind such adjustments is duly justified in Sections~\ref{sec: spanish mkt} and \ref{sec: ndres}.
\begin{subequations} \label{cons:balance_IDM}
\begin{multline} \label{cons: energy_bal_idm1}
    \!\!\!\!\!-\!\!\left(
    \sum_{d\in\mathscr{D}} \!
    \left(
    1 \!+\! \bar{P}_{d,t} 
    \right) 
    P_{d,p,t}^{\rm p^*} \!+\!  \!
    \sum_{\theta\in\Theta} \bar{P}_{\theta}^+
    \right) \!\!
    \leq 
    p_t^{\rm DA^*} \!\!+\! 
    \sum_{\kappa=1}^{k-1}p_{\kappa,t}^{\rm ID^*}  \!+
    p_{k,t}^{\rm ID} \\
    \leq
    \sum_{c\in\mathscr{C}} \bar{P}_{c} \!+\! 
    \sum_{r\in\mathscr{R}} \check{P}_{r,t} \!+\!  
    \sum_{\theta\in\Theta} \bar{P}_{\theta} ~,
    \qquad
    \forall k \in \mathscr{K},  
    \forall t \geq \tau \!\! 
\end{multline} \vspace{1ex}
\begin{IEEEeqnarray}{lr}
    p_t^{\rm DA^*} \!\!+\!  
    \sum_{\kappa=1}^{k-1}p_{\kappa,t}^{\rm ID^*} \!+\!  
    p_{k,t}^{\rm ID} = \!\!\!
    \sum_{b\in \mathscr{B}^m} p_{b,t}^m ~,
    & \qquad
    \forall k \in \mathscr{K}, 
    \forall t \geq \tau \qquad \label{cons: energy_bal_idm3}  \vspace{1ex}
\end{IEEEeqnarray} 
\end{subequations} 
In \eqref{cons:balance_IDM}, $p_t^{\rm DA^*}\!\!$ and $p_t^{\rm ID^*}\!\!$ are the solutions of the \ac{dam} and previous \acp{idm} respectively; and ${P}_{d,p,t}^{\rm p^*}$ is the optimal load profile for each demand, scaled by the profile's upper bound of uncertainty, $\bar{P}_{d,t}$, (see \eqref{cons: energy_bal_idm1}). Note that for the nodal equations, we have the same constraints as in \eqref{cons:balance common}, but the time index, $\forall t \in \mathscr{T}$, is replaced with $\forall t \geq \tau$, where $\tau$ is the first delivery period of the current \ac{idm} session.

\subsection{\aclp{ndres}} \label{sec: ndres}
The \acp{ndres} modeled in \eqref{cons:ndres} comprise mainly wind power and solar PV plants. The lower bound represents the asset technical minimum (e.g., cut-in speed for wind plant) while the output is bounded above by the available stochastic source. 
\vspace{-3mm}  
\begin{IEEEeqnarray}{lr} \label{cons:ndres}
    \ubar{P}_{r,t}
    \leq 
    p_{r,t}
    \leq 
    \check{P}_{r,t}~,
    & \qquad \qquad \qquad \qquad
    \forall r \in \mathscr{R}, 
    \forall t \in \mathscr{T} \qquad
\end{IEEEeqnarray}

\subsection{\aclp{stu}}
\acp{stu} are synchronous generating plants that rely on a stochastic renewable source (solar irradiation). Additionally, they usually include storage capability for as much as 8 hours at full power capacity \cite{garcia2011performance}. Moreover, conversion between thermal and electrical power that takes place in the Power Block (PB) of the \ac{stu} needs to be appropriately formulated. For these reasons, \acp{stu} cannot be accurately represented by parametrizing any of the sets of constraints above, and they require a specific set in the \ac{vpp} model proposed in this paper.

Equation~\eqref{cons:stu_psf} defines upper and lower bounds of the active power (thermal) that can be generated by the solar field, $p_{\theta,t}^{{\rm SF}}$, which is only limited by the available power extractable from solar irradiation. The charging and discharging power (thermal) of the \ac{stu} storage device are bounded above and below in~\eqref{cons:stu_pcharge} and~\eqref{cons:stu_pdisch}. 
The power (thermal), $p_{\theta,t}^{{\rm PB}}$, sent to the PB which converts the thermal into electrical power through a synchronous turbine, is given in~\eqref{cons:stu_pb}. It is the sum of the thermal power generated by the solar field, net thermal power of the \ac{stu} storage and a final factor containing coefficient $K_{\theta}$ that takes startup losses into account. This power is then bounded in~\eqref{cons:stu_pb_bounds} by the maximum and minimum power output of the turbine. The commitment status of the PB reflecting its on/off status is modeled like dispatchable power plants.

\begin{figure} [t!]
    \centering
    \vspace*{0mm}
    \includegraphics[width=\linewidth]{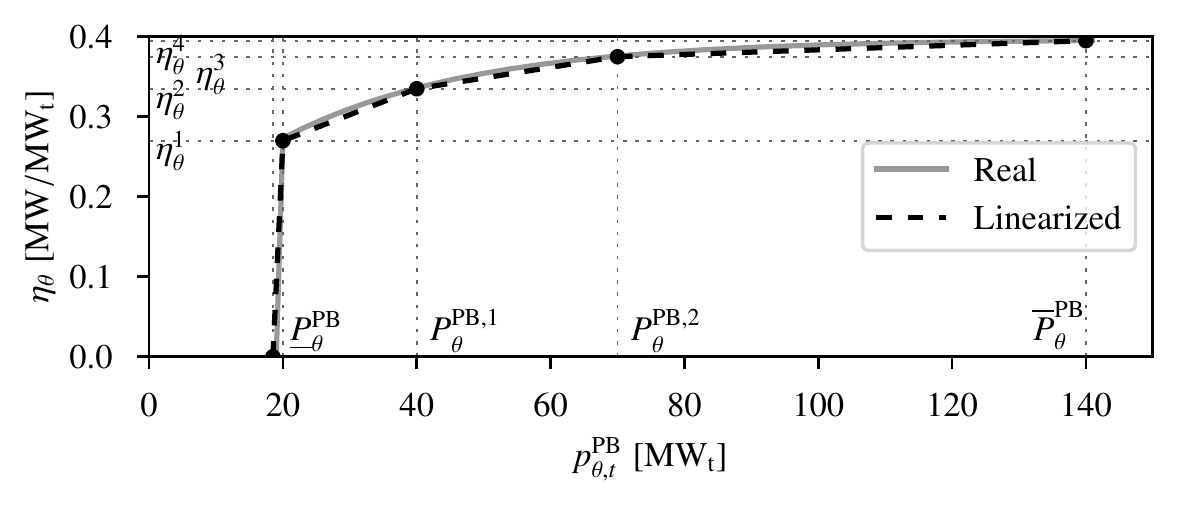}
    \caption{Power Block conversion efficiency}
    \vspace*{-1mm}
    \label{fig:PB_Conv_Eff}
\end{figure}

Electrical power output of the \ac{stu} is given by~\eqref{cons:stu_output}. Conversion between thermal and electrical power is nonlinear, and depends on the level of thermal power injected into the PB. The higher $p_{\theta, t}^{{\rm PB}}$ is, the more efficient the conversion. In this work, we have defined four linear segments, delimited by $P_{\theta}^{\rm PB} \!=\! \{\ubar{P}_{\theta}^{\rm PB}, P_{\theta}^{\rm PB,1}, P_{\theta}^{\rm PB,2}, \bar{P}_{\theta}^{\rm PB} \}$, and characterized by different conversion factors, $\eta_{\theta} \!=\! \{\eta_{\theta}^{1} , \eta_{\theta}^{2} , \eta_{\theta}^{3}, \eta_{\theta}^{4}\}$. Fig.~\ref{fig:PB_Conv_Eff} shows the PB efficiency for converting thermal power input into electric power using parameters for a $50$ MW \ac{stu} in Spain.
\vspace{0.5em}

Finally, equations that model the thermal energy stored in the \ac{stu} storage device are~\eqref{cons:stu_energy_balance}-\eqref{cons:stu_alpha} where $\bar{E}_{\theta,t}$ is the thermal storage capacity and $\ubar\alpha_\theta/\bar\alpha_\theta$ is the lower/upper bound multiplier of storage at the final period of the schedule. \vspace{1ex}
\begin{subequations} \label{cons:stu}
\begin{IEEEeqnarray}{lr} 
    0 
    \leq 
    p_{\theta,t}^{{\rm SF}} 
    \leq 
    \check{P}_{\theta,t} ~,
    &
    \forall \theta \in \Theta, 
    \forall t \in \mathscr{T} \qquad \label{cons:stu_psf} \\
    \ubar{P}_{{\theta}}^ + 
    u_{\theta,t}^+ 
    \leq 
    p_{\theta,t}^ + 
    \leq 
    \check{P}_{\theta,t} 
    u_{\theta,t}^+~,
    &
    \forall \theta \in \Theta, 
    \forall t \in \mathscr{T} \qquad \label{cons:stu_pcharge}\\
    (1 - 
    u_{\theta,t}^+)\ubar{P}^-_{{\theta}} 
    \leq 
    p_{\theta,t}^- 
    \leq 
    (1 - u_{\theta,t}^+)\bar{P}^-_{\theta} ~,
    &
    \forall \theta \in \Theta, 
    \forall t \in \mathscr{T} \qquad \label{cons:stu_pdisch}\\
    p_{\theta, t}^{{\rm PB}} = 
    p_{\theta,t}^{{\rm SF}} + 
    p_{\theta,t}^- - 
    p_{\theta,t}^+ -
    K_\theta v_{\theta,t}^{1} \bar{P}_{\theta}^{{\rm PB}} ~,
    &
    \forall \theta \in \Theta, 
    \forall t \in \mathscr{T} \qquad \label{cons:stu_pb}\\
    u_{\theta, t} \ubar{P}_{\theta}^{{\rm PB}} 
    \leq
    p_{\theta, t}^{\rm PB} 
    \leq 
    u_{\theta,t} \bar{P}_{\theta}^{{\rm PB}} ~,
    &
    \forall \theta \in \Theta, 
    \forall t \in \mathscr{T} \qquad \label{cons:stu_pb_bounds} 
\end{IEEEeqnarray} 
\begin{multline}
    \!\!\!\!\!
    p_{\theta,t} = 
    \left\{
    \setlength\arraycolsep{2pt}
      \begin{matrix}
        \eta_{\theta}^{1} p_{\theta,t}^{{\rm PB}} & 
         \!\!\!\!\!{{\rm if}} \!\!\!\!\! & 
         \quad 
         0 
         \leq 
         p_{\theta,t}^{{\rm PB}} <
         {\ubar{P}_{\theta}^{{\rm PB}}} \\ 
        \eta_{\theta}^{2} p_{\theta,t}^{{\rm PB}} & 
        {{\rm if}} & \;
        \ubar{P}_{\theta}^{{\rm PB}} 
        \leq 
        p_{\theta,t}^{{\rm PB}} < 
        P_{\theta}^{\rm PB,1}  \\ 
        \eta_{\theta}^{3} p_{\theta,t}^{{\rm PB}} & 
        {{\rm if}} &  
        P_{\theta}^{\rm PB,1} 
        \leq 
        p_{\theta,t}^{{\rm PB}} <  
        P_{\theta}^{\rm PB,2} \\ 
        \eta_{\theta}^{4}\, p_{\theta,t}^{{\rm PB}} & 
        {{\rm if}} & \! 
        P_{\theta}^{\rm PB,2} 
        \leq 
        p_{\theta,t}^{{\rm PB}} 
        \leq 
        \bar{P}_{\theta}^{{\rm PB}}
      \end{matrix} 
    \right\}~, \\
    \qquad \qquad \qquad \qquad \qquad \qquad \qquad \quad
    \forall \theta \in \Theta, 
    \forall t \in \mathscr{T} \! \label{cons:stu_output}
\end{multline}
\begin{IEEEeqnarray}{lr} 
    e_{\theta,t} \!=\! 
    e_{\theta, (t-1)} \!+\!
    p_{\theta,t}^+ \Delta t \eta_\theta^+ \!-\!
    \dfrac{p_{\theta,t}^- \Delta t}{\eta_\theta^-}~,
    \phantom{p_{\theta,t}^ \Delta}
    & 
    \forall \theta \in \Theta, 
    \forall t \in \mathscr{T} \qquad \label{cons:stu_energy_balance}  \\
    \ubar{E}_{\theta,t} 
    \leq
    e_{\theta,t} 
    \leq
    \bar{E}_{\theta,t}~, \phantom{\forall \theta \in \Theta, \theta}
    &
    \forall \theta \in \Theta, 
    \forall t \in \mathscr{T} \qquad \label{cons:stu_energy_bounds} \\
    \ubar\alpha_\theta\bar{E}_{\theta,t} 
    \leq 
    e_{\theta,t} 
    \leq 
    \bar\alpha_\theta\bar{E}_{\theta,t}~, 
    &
    \forall \theta \in \Theta, \,\,t = T \qquad \label{cons:stu_bounds_last_period} \\
    0 
    \leq 
    \ubar \alpha_\theta 
    \leq 
    \bar \alpha_\theta \leq 1~,
    & 
    \forall \theta \in \Theta
    \phantom{,\forall t \in \mathscr{T}} \qquad \label{cons:stu_alpha} \\
    u_{\theta,t}, 
    u_{\theta,t}^{+}, 
    v_{\theta,t}^{1} 
    \in 
    \{0,1\}~,
    & 
    \forall \theta \in \Theta, 
    \forall t \in \mathscr{T} \qquad \vspace{1ex}
\end{IEEEeqnarray}
\end{subequations}
Note that, for implementation purposes, the piece-wise linear function, \eqref{cons:stu_output}, is remodelled with SOS-2 constraints \cite{aimms2018aimms}. 

\subsection{Flexible Demands}
In this work, we present a novel participation of flexible demands in a \ac{vpp} comprising two levels of flexibility which are associated with the \ac{dam} and \ac{idm} market sessions. 

\subsubsection{DAM Formulation}
At the first stage, such participation involves the selection of a specific load profile. With this aim, \eqref{cons: demand profiles} and \eqref{cons: demand choice} ensure that for each demand $d$, only one load profile $p$ out of all available profiles is selected. These profiles mirror the different available operational characteristics of the demand assets. e.g. for residential demand, a profile might feature dual peaks at 09:00 and 20:00 while another profile features a shift of the peaks to 07:00 and 21:00 respectively. These profiles are prepared by the demand owners/aggregators and subsequently communicated to the VPP manager. One of them could be the default profile which the load owner will follow if no coordination exists. Other profiles which can offer flexibility but might lead to higher operation costs for the load owner can be presented and the VPP manager needs to compensate them in exchange for the offered flexibility if those profiles are selected.
\begin{subequations} \label{cons:flexload_DAM}
\begin{IEEEeqnarray}{lr}
    p_{d,t} = 
    \sum_{p \in \mathscr{P}} 
    P_{d,p,t} u_{d,p} ~,
    & \qquad \qquad \qquad
    \forall d \in \mathscr{D}, 
    \forall t \in \mathscr{T} \qquad \; \label{cons: demand profiles} \\
    \sum_{p \in \mathscr{P}} u_{d,p} = 
    1 ~,
    & 
    \qquad
    \forall d \in \mathscr{D} 
    \phantom{,\forall p \in \mathscr{P}} \qquad \label{cons: demand choice} \\
    u_{d,p}
    \in 
    \{0,1\}~,
    & 
    \qquad
    \forall d \in \mathscr{D}, 
    \forall p \in \mathscr{P} \qquad \label{cons: demand binary var}
\end{IEEEeqnarray} \vspace{-2mm}
\end{subequations}

\subsubsection{IDM Formulation}
The second level of demand flexibility is provided during the different \ac{idm} sessions, formulated in \eqref{cons:flexload_IDM}.  At \ac{idm}, the load profile selected from \ac{dam} cannot be changed. However, the demand owner allows the VPP manager to vary the consumption a small percentage above and below that selected profile ($P_{d,p,t}^*$) as presented in \eqref{cons: load flex}. These bounds on the demand may or may not be symmetric; and are chosen in this paper as a reflection of realistic practises. Equations~\eqref{cons: load ramp1} and \eqref{cons: load ramp2} define the ramps of the demand profile from one period to the next. Finally, \eqref{cons: load minimum} ensures that, over the total duration of the current \ac{idm} session plus the periods covered in previous sessions, a minimum amount of energy is consumed. The energy values settled in previous periods, $p_{d,t}^*$, are thus accounted for during subsequent \ac{idm}.
\begin{subequations} \label{cons:flexload_IDM}
\begin{multline} \label{cons: load flex}
    \left(
    1 - \ubar{P}_{d,t}
    \right) 
    P_{d,p,t}^{*}
    \leq p_{d,t} 
    \\
    \qquad \qquad \leq   
    \left(
    1 + \bar{P}_{d,t} 
    \right) 
    P_{d,p,t}^{*}~,\!
    \qquad 
    \forall d \in \mathscr{D}, 
    \forall t \geq \tau \,\, \hspace{-0.6em}
\end{multline}
\begin{IEEEeqnarray}{lr}
    p_{d,t} - 
    p_{d,(t-1)} 
    \leq 
    \bar{R}_d \Delta t~,
    & 
    \forall d \in \mathscr{D}, 
    \forall t \geq \tau \quad \label{cons: load ramp1} \\
    p_{d,(t-1)} - 
    p_{d,t} 
    \leq 
    \ubar{R}_d \Delta t~,
    & 
    \forall d \in \mathscr{D}, 
    \forall t \geq \tau \quad \label{cons: load ramp2} \\
    \ubar{E}_d 
    \leq 
    \sum_{t=1}^{\tau-1} p_{d,t}^* \Delta t + 
    \sum_{t= \tau}^{\left| \mathscr{T} \right|} p_{d,t} \Delta t~,
    & \qquad
    \forall d \in \mathscr{D} 
    \phantom{,\forall t \geq \tau} \quad \label{cons: load minimum}
\end{IEEEeqnarray}
\end{subequations}

\section{Case Study} \label{sec: case study}
This section presents the case studies considered to test and validate the RES-based \ac{vpp} model proposed in this paper. Section IV.A outlines the \ac{vpp} topology considered, which resembles a subarea of the southern region of the Spanish grid. The input data that is fed into the model is then described in Section IV.B. Finally, Section IV.C presents and discusses the simulation results. These case studies are implemented in GAMS with CPLEX on a computer with 32GB memory, intel i7 and run-time (reading of data, algorithm execution and writing of results) is less than 5 seconds for all scenarios.

\subsection{VPP Description}
\begin{figure}[t!]
    \centering
    \vspace*{0mm}
    \includegraphics[width=\linewidth, height=1.6in]{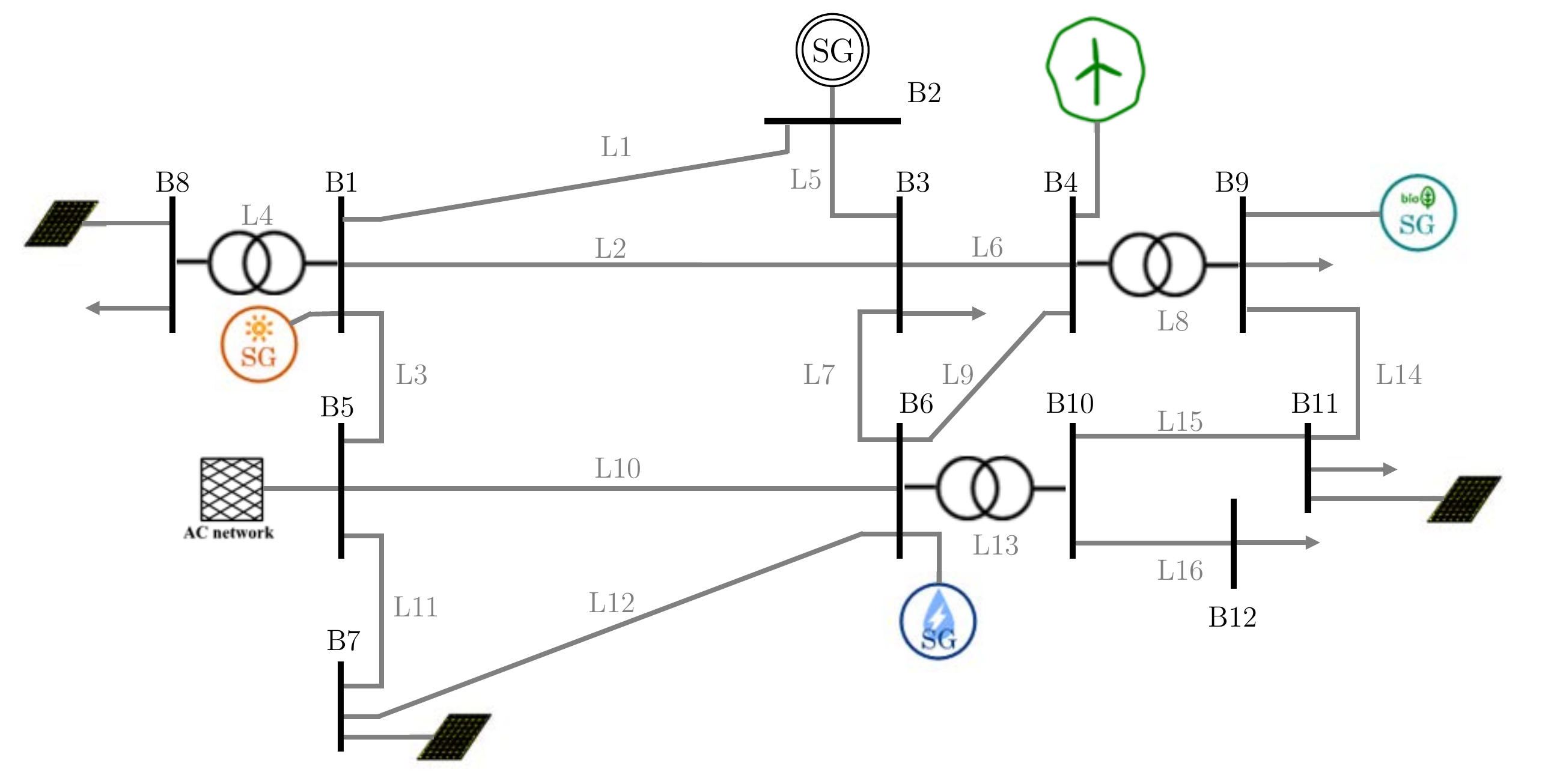}
    \caption{12-node network for test cases}
    \vspace*{0mm}
    \label{fig:network structure}
\end{figure}

The \ac{vpp} assets are distributed across a 12-node network connected to a main grid through a \ac{pcc} as shown in Fig.~\ref{fig:network structure}. 
\color{black}
The capacities of \acp{dres}, hydro (bus 6) and biomass (bus 9) are $111$ and $5$ MW respectively. For \acp{ndres}, wind power plant (bus 4), solar PV (bus 8) and \ac{stu} (bus 1) power block have rated capacities of $50$ MW each. The thermal storage capacity associated with the \ac{stu} is $1100$ MWh$_{\rm th}$. It is desired that, after an operation day, a predefined amount of such capacity is reserved for first period of the next operation day such that it might be used to capture some benefits of early high prices before another charging period begins (see eqs.~\eqref{cons:stu_bounds_last_period} and~\eqref{cons:stu_alpha}).
Additionally, thermal storage can only be charged from the solar field and not from the main grid. The demands considered are industrial, airport and residential loads (buses 3, 9 and 12)  with minimum daily consumption of $800$, $580$, and $600$ MWh respectively. Three profiles are associated with each demand and total consumption for each profile is same.

\subsection{Input Parameters}
\begin{figure}[b!]
    \centering
    \vspace*{0mm}
    \includegraphics[width=\linewidth]{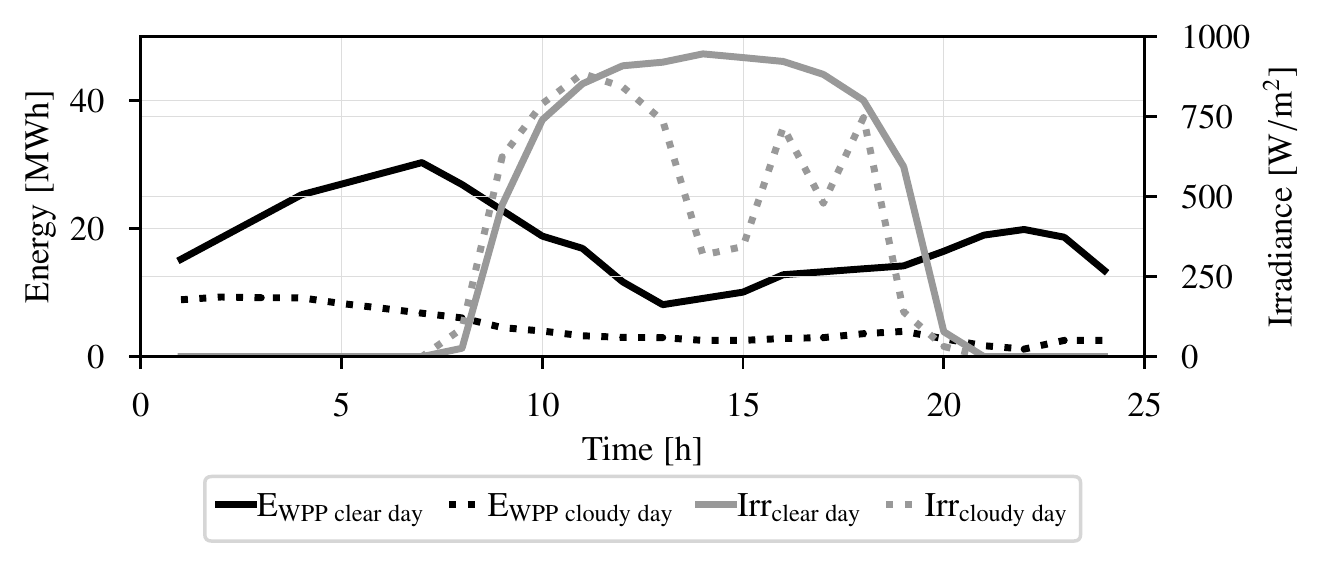}
    \caption{Forecasts for \ac{stu}, WPP \& PV generation on a clear and cloudy day}
    \vspace*{0mm}
    \label{fig:Forecast WPP Irr}
\end{figure}

A time horizon of $24$ hours with an hourly timescale is utilized during \ac{dam}, while a subset of the $24$ hours is used for each \ac{idm} session (see~\cite{chaves2015spanish}). 
The VPP has the possibility to supply its internal demand, partially or entirely, by buying energy from the main grid at the \acp{pcc} if needed. In our case, the maximum value of energy that the VPP can buy is $110$ MWh. 
Furthermore, we make computations for different seasons and days based on historical data of actual and forecast profiles. Fig.~\ref{fig:Forecast WPP Irr} shows estimated wind power output and irradiation levels in Southern Spain for a clear, sunny day in March 2014 and a day with cloud covers in March 2018. 

\begin{figure}[tbh]
    \centering
    \includegraphics[width=\linewidth]{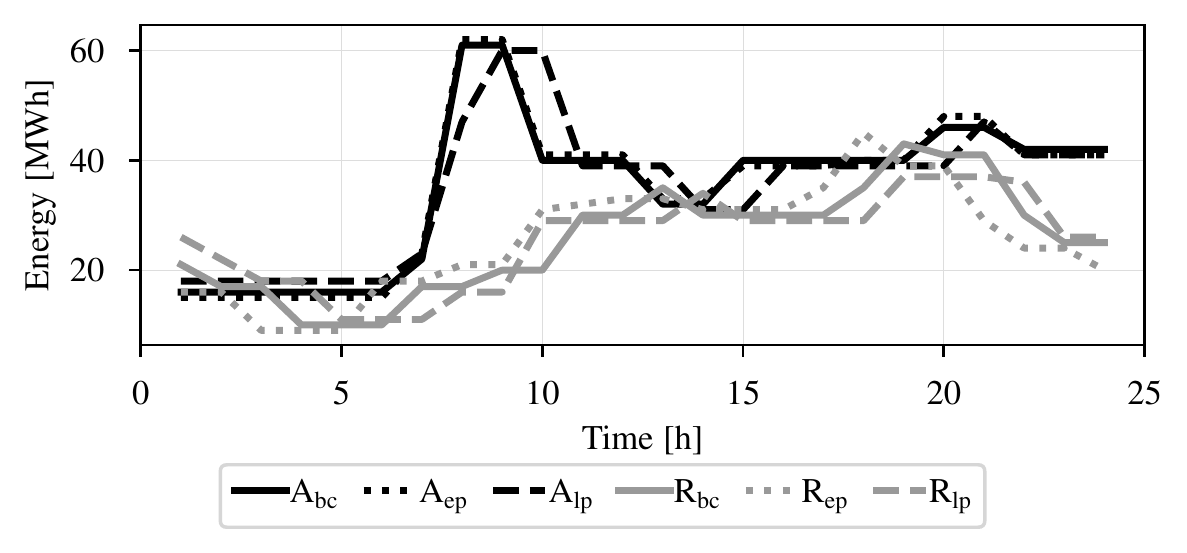}
    \caption{Airport and residential demand profiles}
    \label{fig:load profiles example}
\end{figure}

For demands, we show as example two of the three loads: airport (A) and residential (R) with three different profiles ('bc-basecase', 'ep-early peak', 'lp-late peak') in Fig.~\ref{fig:load profiles example}. 
While fictitious, the base profile replicates \textit{default activities} of airport and residential electricity consumers while other profiles are designed to perform peak shifting around it. During \ac{idm}, the demand owner allows a $10\%$ tolerance for demand movement over the selected profile at \ac{dam}. \color{black} 

\begin{figure}[tbh]
    \centering
    \vspace*{0mm}
    \includegraphics[width=\linewidth]{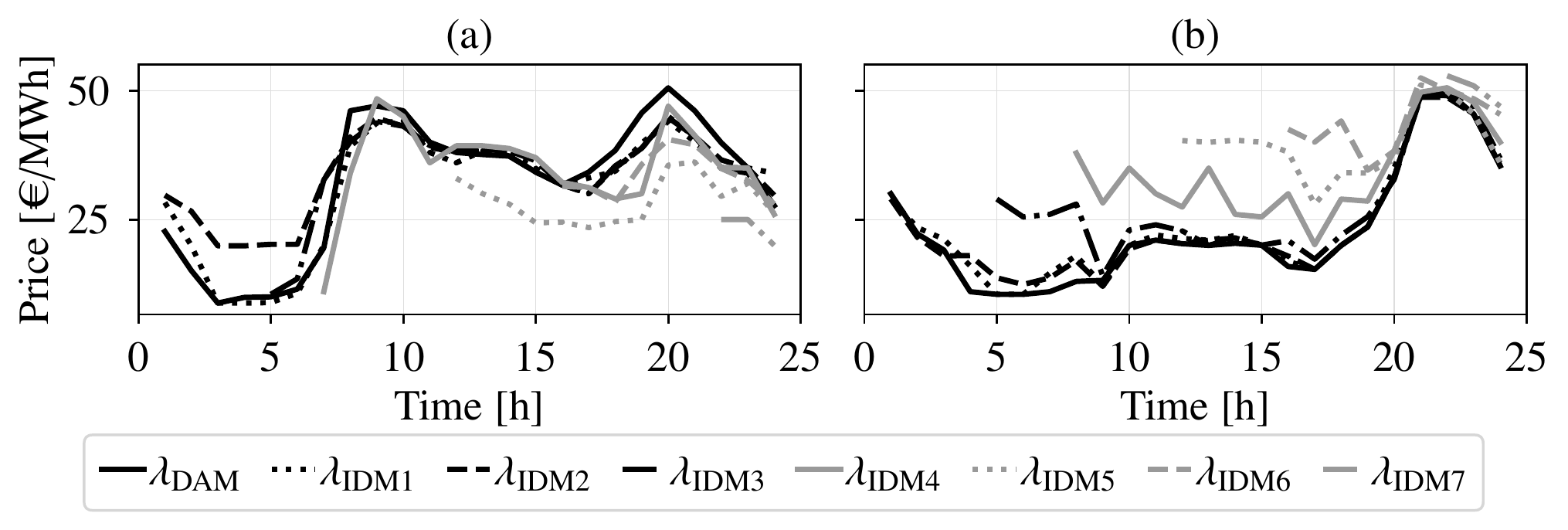}
    \caption{DAM and IDM prices on (a) clear day, (b) cloudy day}
    \vspace*{0mm}
    \label{fig:DAM IDM Prices}
\end{figure}

In Fig~\ref{fig:DAM IDM Prices}, representative prices for \ac{dam} and all \ac{idm} sessions for both clear and cloudy days are presented.

\subsection{Results}
To test the effectiveness of our model, we identify two operation strategies: 
\begin{itemize}
    \item Base case where generation or consumption units act individually (No Coordination). 
    \item Coordinated \ac{vpp} operation as  formulated in Section \ref{sec: vpp model}.
\end{itemize}

\begin{figure}[t!]
     \centering
     \vspace*{0mm}
     \includegraphics[width=\linewidth]{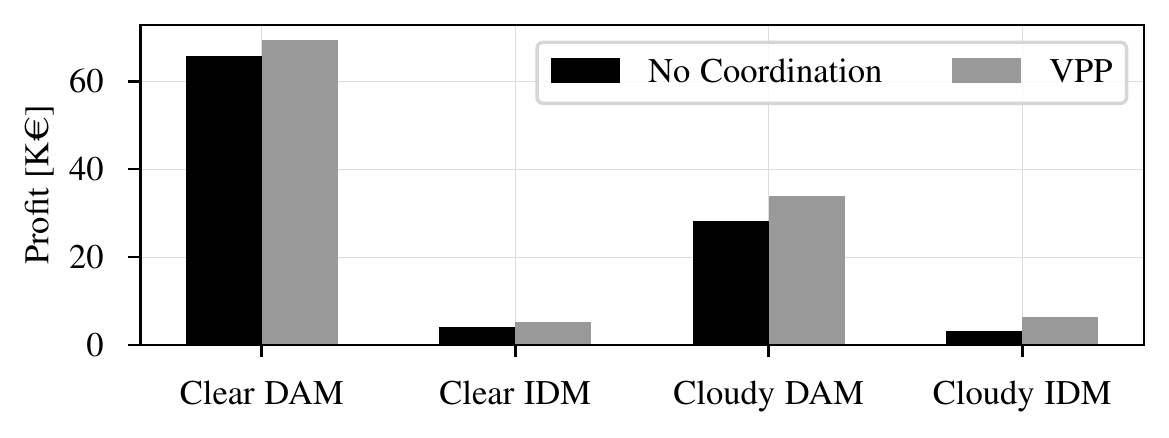}
     \caption{VPP benefits on different strategies}
     \vspace*{0mm}
     \label{fig:VPP benefits}
\end{figure}

First, we outline the benefits of the \ac{vpp} strategic behaviour over the No Coordination case for the \ac{dam} and the first \ac{idm} session with a default load profile in both cases. 
Profit comparison between the days and the strategies is shown in Fig.~\ref{fig:VPP benefits}. For the clear day, there is a $5\%$ and $28\%$ increase in the profits during \ac{dam} and \ac{idm} respectively when the \ac{vpp} model is compared to the case without coordination. However, these profits are more significant during the cloudy day where the \ac{vpp} outperforms the base case with $20\%$ and $99\%$ during \ac{dam} and \ac{idm} respectively. 

To fully comprehend the effectiveness of the presented \ac{vpp} model, we analyse the following scenarios in detail: firstly, a clear day with high wind power plant output, high irradiance level, and secondly, a day with intermittent cloud covers in the afternoon and low wind power output. Each of these two broad categorisations will be further examined under (a) Base case with default load profile (b) \ac{vpp} with default load profile and (c) \ac{vpp} with three load profiles at zero cost. Finally, we conduct an analysis to ascertain the behaviour of non-zero cost demand profiles on a clear day.

\subsubsection{Traded Power and Output of Assets on a Clear Day}
\begin{figure*}[ht]
     \centering
     \vspace*{-2mm}
     \includegraphics[width=1\linewidth]{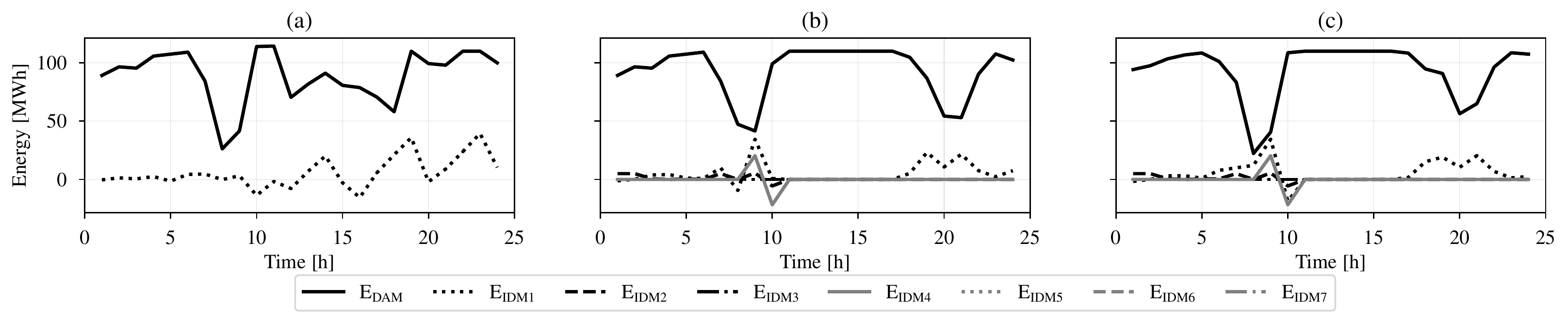}
     \caption{Traded power on a clear day at DAM \& IDM in: (a) No Coordination, (b) VPP with default load profile, (c) VPP with three load profiles at zero cost}
  \label{fig:Traded Power on a Clear Day}
\end{figure*}

\begin{figure*}[ht]
     \centering
     \vspace*{-2mm}
     \includegraphics[width=1\linewidth]{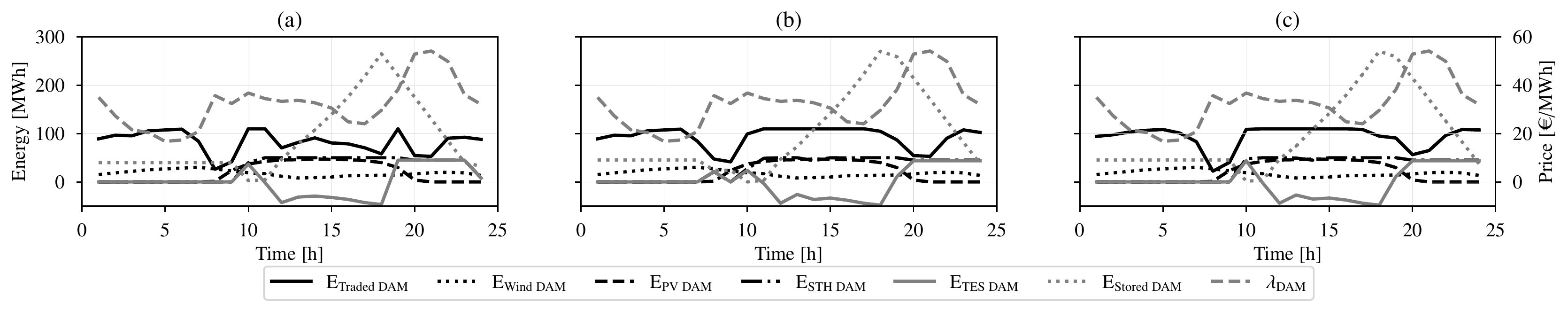}
     \caption{Output of assets at DAM on a clear day: (a) No Coordination, (b) VPP with default load profile, (c) VPP with three load profiles at zero cost}
     \label{fig:Output of Assets at DAM Clear Day}
\end{figure*}

Fig.~\ref{fig:Traded Power on a Clear Day} shows traded power on a clear day during \ac{dam} and all \ac{idm} sessions and Fig.~\ref{fig:Output of Assets at DAM Clear Day} depicts the expected behaviour of the different assets at \ac{dam}. The effect of coordination is evident from the smoothing of the traded power curve from period 10h-16h in \ac{vpp} compared with the base case. The differences arise largely due to proper synchronization of the \ac{stu} with its storage and the biomass plant response to the market price. The low volume of energy traded in period 8h-9h coincide with very high demand and relatively low generation in the three cases.
Moreover, in terms of benefits to the \ac{vpp}, case (c) with three load profiles at zero cost has the highest profit because it leverages differences in the load shape and chooses the best profile to suit its objective. The residential early peak profile was selected in this instance while the basecases were used both for the industrial and airport demand.



For \ac{idm} participation, after updates to prices and output forecasts of \ac{ndres}, the \ac{vpp} manager (or owner of each asset in case of No Coordination) makes adjustments to \ac{dam} auctions and in most cases, prioritizes the \ac{ndres} over the \ac{dres} units. Additionally, most updates are made in the earlier \ac{idm} sessions as evidenced in Fig~\ref{fig:Traded Power on a Clear Day} where there are no changes in offers made in \ac{idm}5 - \ac{idm}7.

\subsubsection{Traded Power and Output of Assets on a Cloudy Day}
\begin{figure*}[t!]
     \centering
     \vspace*{0mm}
     \includegraphics[width=1\linewidth]{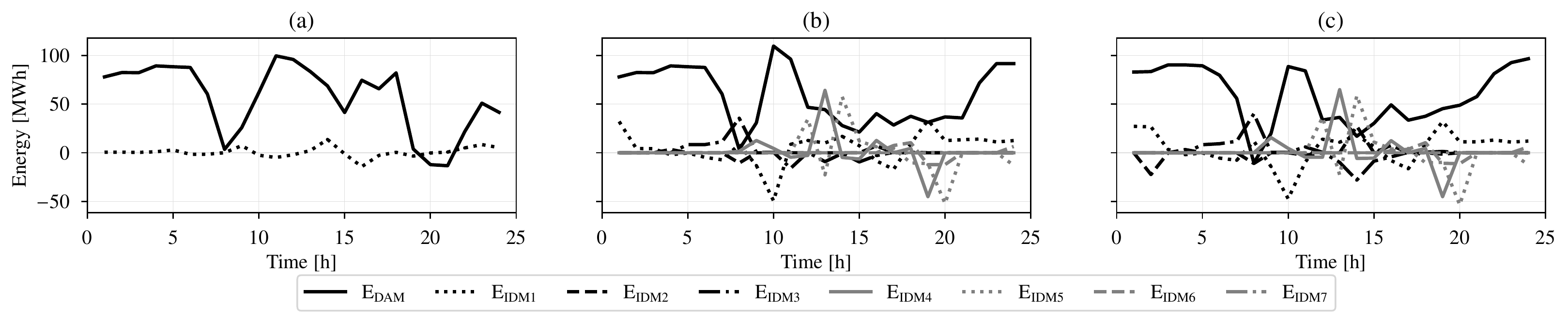}
     \caption{Traded power on a cloudy day at DAM \& IDM: (a) No Coordination, (b) VPP with default load profile, (c) VPP with three load profiles at zero cost}
  \label{fig:Traded Power on a Cloudy Day}
\end{figure*}

The \ac{vpp} behaves differently on a day with intermittent cloud covers in the afternoon. There are more activities during \ac{idm} as the forecasts of \ac{ndres} are in flux as shown in Fig.~\ref{fig:Forecast WPP Irr}. This uncertainties, however, are taken care of by the model as observed in Fig.~\ref{fig:Traded Power on a Cloudy Day} (b) and Fig.~\ref{fig:Traded Power on a Cloudy Day} (c) where \ac{idm} bids reflect their expected behavior. With respect to demand, the early peak profiles were selected for both industrial and residential demands whereas the base case was retained airport demand.

\begin{figure*}[ht]
     \centering
     \vspace*{-2mm}
     \includegraphics[width=1\linewidth]{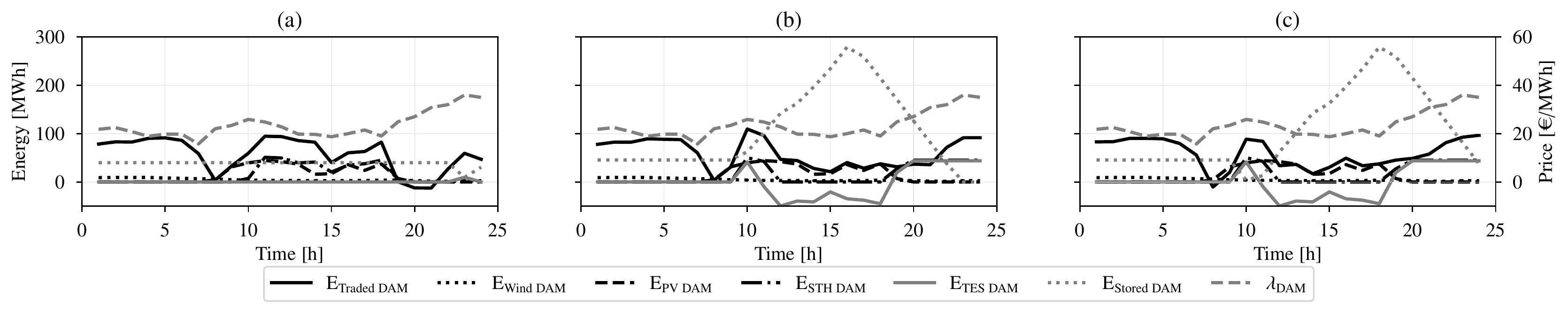}
     \caption{Output of assets at DAM on a cloudy day: (a) No Coordination, (b) VPP with default load profile, (c) VPP with three load profiles at zero cost}
     \label{fig:Output of Assets at DAM Cloudy Day}
\end{figure*}

The comparative advantage of the \ac{vpp} over base case was the operation of the \ac{stu} at the early hours. The \ac{stu} charged the thermal storage instead of delivering energy from solar field to PB at lower efficiency, and thereafter delivered it at high PB efficiency while also capturing the high prices towards the end of the day as shown in Fig.~\ref{fig:Output of Assets at DAM Cloudy Day} (b) and Fig.~\ref{fig:Output of Assets at DAM Cloudy Day} (c). 

\subsubsection{Effects of Non-zero Cost on Demand Profile Choice}
Here, we provide a solution to the incentive structure for demand owners. The default load profile has zero cost because the demand owner follows it irrespective of other events. The \ac{vpp} then decides what price to pay the demand owners for other profiles such that the \ac{vpp}'s benefits are not eroded. 

In Table~\ref{table:demand profile choice}, the optimal choices of demand profiles are shown. Industrial and residential loads have the early peak and late peak as optimal profiles respectively. However, \ac{vpp} manager is only willing to pay up to \euro$320$/day to the industrial load owner after which it is no more profitable for the \ac{vpp} operation. In both of these demand cases, the industrial late peak profile is not profitable at all for the \ac{vpp} and neither is the residential early peak profile. In the case of the airport load, there are more options for both the airport demand owner and the \ac{vpp} manager. The early peak profile is profitable for the \ac{vpp} manager as a cost up until \euro$500$/day while the late peak profile is only profitable until \euro$305$/day.

\begin{table}[ht]
  \centering
  \caption{Demand profile choice at non-zero costs}
 \vspace{1mm}
  \begin{tabular}{llcc}
    \toprule
    \multicolumn{1}{c}{\textbf{Demand}} & \multicolumn{1}{c}{\textbf{Basecase}} &    \multicolumn{1}{c}{\textbf{Early peak}} & \multicolumn{1}{c}{\textbf{Late peak}} \\
    \midrule
    \multirow{1}{*}{Industrial} & chosen when & \multirow{1}{*}{optimal} & \multirow{1}{*}{\xmark} \\
    & cost$>$€320/day &  & \\[0.5em]
    \multirow{1}{*}{Residential} & chosen when & \multirow{1}{*}{\xmark} & \multirow{1}{*}{optimal} \\
    & cost$>$€180/day &  & \\[0.5em]
    \multirow{1}{*}{Airport} & chosen when   & \multirow{1}{*}{optimal} & \multicolumn{1}{l}{suboptimal - chosen}  \\
    & cost$>$€500/day  & & \multicolumn{1}{l}{when cost$>$€305/day}  \\    
    \bottomrule
  \end{tabular}
  \label{table:demand profile choice}
\end{table}

\section{Conclusion} \label{sec: conclusion}
This paper presents a heterogeneous \ac{res}-based \ac{vpp} that participates in energy markets. Coupled with standard models of dispatchable resources, improved formulations of flexible demand and concentrated solar thermal plant were introduced to describe the technical operation of the \ac{vpp} units. A network constrained unit commitment model is thereafter utilized by the \ac{vpp} to participate in market activities. The energy market chosen in our study is the Spanish market. 

Different case studies were then analysed to test the model's robustness against uncertainties due to weather conditions.
\begin{enumerate}
    \item The \ac{vpp} outperforms the case where the units are not coordinated by up to 20\% on cloudy days.
    \item \acp{vpp} incorporating flexible demand profiles with zero costs are the most beneficial compared to other configurations because the model selects the best profile that maximizes its profit.
    \item When cost of demand profiles are non-zero, there are some thresholds which the \ac{vpp} is willing to pay the demand owners until it becomes less profitable and they return to the default profile.
    \item The impact of the \ac{stu} with its storage was shown where the storage was charged at early periods while later discharging at higher capacity with higher efficiency. This led to higher profits of \ac{vpp} over No Coordination in all conditions and especially during the cloudy days.
\end{enumerate}

Future directions for this work include co-optimizing the \ac{dam} planning with secondary reserve market bids. This could ensure that there is available tolerance during real time operation. Application of the model to larger systems and considering AC power flow constraints will also be studied.

\section*{Acknowledgements} \label{sec: ack}
The authors wish to thank Mr. Ione Lopez with Iberdrola, Spain, and the people with CIEMAT, Spain, especially Dr Mario Biencinto and Dr. Loreto Valenzuela, that participated in the EU project POSYTYF, for the discussions on the modeling of solar thermal plants and flexible demands, and the provision of data used in the case study. 

\bibliographystyle{IEEEtran}
\bibliography{refs}

\begin{thebibliography}{10}
\providecommand{\url}[1]{#1}
\csname url@samestyle\endcsname
\providecommand{\newblock}{\relax}
\providecommand{\bibinfo}[2]{#2}
\providecommand{\BIBentrySTDinterwordspacing}{\spaceskip=0pt\relax}
\providecommand{\BIBentryALTinterwordstretchfactor}{4}
\providecommand{\BIBentryALTinterwordspacing}{\spaceskip=\fontdimen2\font plus
\BIBentryALTinterwordstretchfactor\fontdimen3\font minus
  \fontdimen4\font\relax}
\providecommand{\BIBforeignlanguage}[2]{{%
\expandafter\ifx\csname l@#1\endcsname\relax
\typeout{** WARNING: IEEEtran.bst: No hyphenation pattern has been}%
\typeout{** loaded for the language `#1'. Using the pattern for}%
\typeout{** the default language instead.}%
\else
\language=\csname l@#1\endcsname
\fi
#2}}
\providecommand{\BIBdecl}{\relax}
\BIBdecl

\bibitem{pinson2017towards}
P.~Pinson, L.~Mitridati, C.~Ordoudis, and J.~{\O}stergaard, ``{Towards fully
  renewable energy systems: Experience and trends in Denmark},'' \emph{CSEE
  journal of power and energy systems}, vol.~3, no.~1, pp. 26--35, 2017.

\bibitem{koraki2017wind}
D.~Koraki and K.~Strunz, ``{Wind and solar power integration in electricity
  markets and distribution networks through service-centric VPPs},'' \emph{IEEE
  Trans on Power Systems}, vol.~33, no.~1, pp. 473--485, 2017.

\bibitem{dominguez2014operation}
R.~Dom{\'\i}nguez, A.~Conejo, and M.~Carri{\'o}n, ``{Operation of a fully
  renewable electric energy system with CSP plants},'' \emph{Applied energy},
  vol. 119, pp. 417--430, 2014.

\bibitem{sioshansi2010value}
R.~Sioshansi and P.~Denholm, ``The value of concentrating solar power and
  thermal energy storage,'' \emph{IEEE Trans on Sus Energy}, vol.~1, no.~3, pp.
  173--183, 2010.

\bibitem{He2016}
G.~He, Q.~Chen, C.~Kang, and Q.~Xia, ``Optimal offering strategy for {CSPs} in
  joint energy, reserve and regulation markets,'' \emph{IEEE Trans on Sus
  Energy}, vol.~7, no.~3, pp. 1245--1254, 2016.

\bibitem{rohani2017modelling}
S.~Rohani, T.~Fluri, F.~Dinter, and P.~Nitz, ``Modelling and simulation of
  parabolic trough plants based on real operating data,'' \emph{Solar Energy},
  vol. 158, pp. 845--860, 2017.

\bibitem{liu2018optimal}
Z.~Liu, W.~Zheng, F.~Qi, L.~Wang, B.~Zou, F.~Wen, and Y.~Xue, ``Optimal
  dispatch of a virtual power plant considering demand response and carbon
  trading,'' \emph{Energies}, vol.~11, no.~6, p. 1488, 2018.

\bibitem{de2018local}
{A. De La Nieta et al}, ``Local economic dispatch with local renewable
  generation and flexible load management,'' in \emph{2018 SEST
  Conference}.\hskip 1em plus 0.5em minus 0.4em\relax IEEE, 2018, pp. 1--6.

\bibitem{herter2007residential}
K.~Herter, ``{Residential implementation of critical-peak pricing of
  electricity},'' \emph{Energy policy}, vol.~35, no.~4, pp. 2121--2130, 2007.

\bibitem{hungerford2019value}
Z.~Hungerford, A.~Bruce, and I.~MacGill, ``{The value of flexible load in power
  systems with high renewable energy penetration},'' \emph{Energy}, vol. 188,
  p. 115960, 2019.

\bibitem{ciarreta2011renewable}
A.~Ciarreta, C.~Guti{\'e}rrez-Hita, and S.~Nasirov, ``Renewable energy sources
  in the {Spanish} electricity market: Instruments and effects,'' \emph{Ren.
  and Sus. Energy Reviews}, vol.~15, no.~5, pp. 2510--2519, 2011.

\bibitem{omiees}
``Operador del mercado ibérico de energía,'' https://www.omie.es/en/.

\bibitem{redelectrica}
``Red eléctrica de españa {(REE)},'' https://www.ree.es/en/.

\bibitem{chaves2015spanish}
J.~P. Chaves-{\'A}vila and C.~Fernandes, ``{The Spanish intraday market design:
  A successful solution to balance renewable generation?}'' \emph{Renewable
  Energy}, vol.~74, pp. 422--432, 2015.

\bibitem{baringo2020virtual}
L.~Baringo and M.~Rahimiyan, ``Virtual power plants and electricity markets,''
  in \emph{e-Book}.\hskip 1em plus 0.5em minus 0.4em\relax Springer, 2020.

\bibitem{kargarian2016toward}
{A Kargarian et al}, ``Toward distributed/decentralized {DC} optimal power flow
  implementation in future electric power systems,'' \emph{IEEE Trans on SG},
  vol.~9, no.~4, pp. 2574--2594, 2016.

\bibitem{van2014dc}
K.~Van~den Bergh, E.~Delarue, and W.~D’Haeseleer, ``{DC power flow in unit
  commitment models},'' \emph{Energy and Environment}, vol. 240, 2014.

\bibitem{garcia2011performance}
I.~L. Garc{\'\i}a, J.~L. {\'A}lvarez, and D.~Blanco, ``{Performance model for
  parabolic trough STU with thermal storage: Comparison to operating plant
  data},'' \emph{Solar Energy}, vol.~85, no.~10, pp. 2443--2460, 2011.

\bibitem{aimms2018aimms}
B.~AIMMS, ``Aimms modeling guide—integer programming tricks,'' \emph{Haarlem,
  The Netherlands: AIMMS BV}, 2018.

\end{thebibliography}

\end{document}